\documentclass[conference]{IEEEtran}

\usepackage{graphicx}
\usepackage[normalem]{ulem}
\usepackage{adjustbox}
\usepackage{hhline}
\usepackage{multirow}
\usepackage{url}
\usepackage{rotating}
\usepackage[cmex10]{amsmath}
\usepackage{balance}
\usepackage[backend=bibtex,firstinits=true,style=numeric-comp,sortcites=true,doi=false,isbn=false,url=false,maxnames=99]{biblatex}

\usepackage{fancybox}
\newcommand{\hypobox}[1]{
\begin{center}        \noindent\thicklines\setlength{\fboxsep}{8pt}        \cornersize{0.2}\Ovalbox{
        \begin{minipage}{3in}        \textit{#1}
        \end{minipage}} 
\end{center}}

\newcommand{\RQA}{To what extent the second-order mutants can be used in place of the first-order mutants for estimation purposes?}
\newcommand{\RQone}{How accurately can our model estimate the first-order mutation coverage?}

\newcommand{\RQtwo}{Is the modeled behavior of second-order mutants consistent with the empirical data?}
\newcommand{\RQthree}{How does incrementing the order of mutation affect the accuracy of the estimation provided by the model?}

\begin{document}
\title{A Model to Estimate First-Order Mutation Coverage from Higher-Order Mutation Coverage}

\author{
\IEEEauthorblockN{Ali Parsai, Alessandro Murgia, and Serge Demeyer}
\IEEEauthorblockA{Faculty of Science\\
University of Antwerp\\
Middelheimlaan 1\\
2020 Antwerpen, Belgium\\
Email: \{ali.parsai, alessandro.murgia, serge.demeyer\}@uantwerpen.be}
}

\maketitle

\thispagestyle{plain}
\pagestyle{plain}

\begin{abstract}

The test suite is essential for fault detection during software development. First-order mutation coverage is an accurate metric to quantify the quality of the test suite. However, it is computationally expensive. Hence, the adoption of this metric is limited. In this study, we address this issue by proposing a realistic model able to estimate first-order mutation coverage using only higher-order mutation coverage. Our study shows how the estimation evolves along with the order of mutation. 
We validate the model with an empirical study based on 17 open-source projects. 

\end{abstract}


\section{Introduction}
The advent of agile processes with their emphasis on test-driven development~\cite{Beck2002} and continuous integration~\cite{Fowler2006} implies that developers want (and need) to test their newly changed or modified classes or components early and often~\cite{McGregor2007}. Therefore, the quality of the test suite is an important factor during the evolution of the software. One of the extensively studied methods to improve the quality of a test suite is mutation testing~\cite{DeMillo1978}.  
Mutation testing provides a repeatable and scientific approach to measure the quality of the test suite, and it is demonstrated to simulate the faults realistically~\cite{Andrews2005,Just2014}. This is because the faults introduced by each mutant are modeled after the common mistakes developers often make~\cite{Jia2011}. 
Although the idea of mutation testing has been introduced in the late 1970s~\cite{Hamlet1977,DeMillo1978}, it has not found widespread use in industry due to its computationally expensive nature.  Therefore, several approaches have been proposed in order to make this technique feasible in industrial settings~\cite{Offutt2001}. Exploiting higher-order mutants instead of first-order mutants is one way used in literature to approach this problem~\cite{Jia2011}. Higher-order mutants can be created by partitioning the set of first-order mutants randomly, and combining first-order mutants in each partition into higher-order mutants.
The benefits of higher-order mutation are twofold. First, 
higher-order mutants are less likely to generate false positives than first-order mutants~\cite{Offutt1992,Jia2009}. Equivalent mutants act as false positives~\cite{Grun2009}, and their detection is an undecidable problem~\cite{Nguyen2014}. However, when an equivalent mutant is combined  with a non-equivalent mutant, the resulting higher-order mutant is non-equivalent as well~\cite{Kintis2012}. This means that second-order mutants are  less likely to suffer from the equivalency problem~\cite{Jia2008}.
Second, by using higher-order mutants, fewer mutants need to be evaluated~\cite{Jia2008}. 
For instance, combining first-order mutants two-by-two into second-order mutants would reduce the number of mutants to 50\%. As a consequence, it saves close to half of the computational time and make more feasible the integration of mutation testing in  continuous integration systems.

Despite the benefits, the adoption of higher-order mutants presents also drawbacks and limitations.
When higher-order mutants are constructed in the aforementioned manner, the mutation coverage calculated using higher-order mutants is not as precise as the 
 one calculated using first-order mutants, with the former often overestimating the capabilities of a test suite. While using higher-order mutants allows us to evaluate less mutants, it also means that we lose information regarding  the status of each underlying first-order mutant. This creates a limitation on the order of mutation, as the value of the lost information overcomes the value of the information at hand. 

In this paper, we address these shortcomings. 
We propose a model able to estimate first-order mutation coverage requiring only the computation of higher-order mutation coverage and
 yet with a smaller chance of equivalency problem. 
 We are interested in a \textit{realistic} model, namely a model that explains in a reasonable 
 manner how the first-order mutants coverage affects the higher-order mutant coverage\footnote{Several polynomial functions of order N  can fit the empirical data. 
 Yet, the underlying  model may not be suitable to explain the behavior of the data.}.
 Initially, we evaluate the estimation of our model for the second-order mutants. Then, we extend the analysis to find out whether our model can be used with mutants of third or higher order.   
For each higher-order mutant, we also verify whether our model correctly describes their real behavior.
More specifically, we look for side effects  (e.g.; fault shifting and fault masking ~\cite{Jia2009}) due to the combination of first-order mutants into higher-order mutants.

This work follows the Goal-Question-Metric paradigm~\cite{Basili1993}. 
The \textit{goal} of this study is to build a model based on higher-order mutation coverage able to estimate the first-order mutation coverage. 
The behavior of the empirical data must be explained reasonably by such model.
The \textit{focus} is to explore the accuracy and the limitations of the model according to the order of mutation.  
The \textit{viewpoint} is that of the team leaders and testing researchers, both interested in making mutation testing applicable in real case scenarios.
The \textit{environment} of this study consists of  17 open-source cases. For this reason, we pursue the following research questions: 
\begin{itemize}
	\item \textbf{RQ1.} \RQA
	\item \textbf{RQ2.} \RQthree
	\end{itemize}

The rest of the paper is structured as follows: Section~\ref{sect:background} provides necessary background information about the subject of the study. Section~\ref{sect:model} describes our proposed model. In Section~\ref{sect:csdesign} we explain the design of our case study. In Section~\ref{sect:results} we present  the results of our experiment. Section~\ref{sect:threats} contains the threats to the validity of our study, and our attempts to reduce them. Section~\ref{sect:rw} discusses the related studies, and their differences with ours. Finally, we conclude the paper in Section~\ref{sect:conclusion}.

\section{Background}
\label{sect:background}
In this section we present an overview of the background information needed to understand our proposed model. First we explain the mutation testing, then we discuss the equivalent mutants, and finally, we take a closer look at the higher-order mutants.
 
\subsection*{Mutation Testing}
Mutation testing is the process of injecting faults into software, and counting the number of these faults that make at least one test fail. The idea of mutation testing was first mentioned in a class paper by Lipton
and later developed by DeMillo, Lipton, and Sayward~\cite{DeMillo1978}. The first implementation of a mutation testing tool was done by Timothy Budd in 1980~\cite{Budd1980}.

The procedure for mutation testing is as follows: 
First, faulty versions of the software are created by introducing a single fault into the system\textit{ (Mutation)}. This is done by applying a known transformation on a certain part of the code \textit{(Mutation Operator)}. After generating the faulty versions of the software \textit{(Mutants)}, the test suite is executed on each one of these mutants. If there is an error or failure during the execution of the test suite, the mutant is regarded as killed \textit{(Killed Mutant)}. On the other hand, if all tests pass, it means that the test suite could not catch the fault and the mutant has survived \textit{(Survived Mutant)}.  The final result is calculated by dividing the number of killed mutants by the number of all non-equivalent mutants. This metric provides a detailed assessment of the quality of a test suite, as it makes sure that the kind of faults simulated by the mutation operators are covered by the tests and therefore reduces the chance of missing such faults in the final product.
In addition of using mutation coverage, test developers can target  surviving mutants. This allows killing mutants to be used as a test requirement.

\subsection*{Equivalent Mutants}

\begin{figure}
	\centering
	\includegraphics[width=0.5\linewidth]{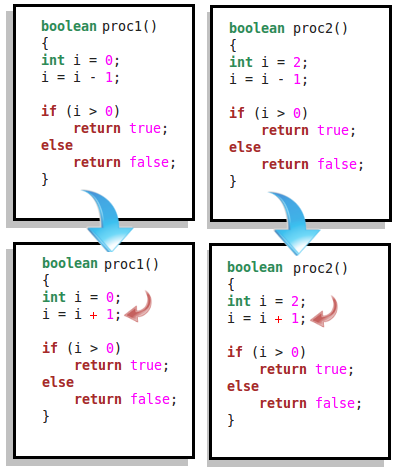}
	\caption{Example for emergence of equivalence due to the context}
	\label{fig:equivalent-context}
\end{figure}

If a mutant produces the same output as the original program for all input values, it is called an equivalent mutant. The creation of equivalent mutants is undesirable~\cite{Grun2009}, but they are not easy to detect~\cite{Madeyski2014}. The creation of such mutants depends on the context of the program itself, for example, in Figure~\ref{fig:equivalent-context}, the introduced change in \textit{proc1} will change the outcome, while the same change in \textit{proc2} does not. The error introduced in \textit{proc2} is, therefore, not detectable by any test. The existence of equivalent mutants would create false positives, since there is no way for a test suite to catch them.

\subsection*{First-Order Mutants and Higher-Order Mutants} 
First-order mutants are the mutants generated by applying a mutation operator on the source code only once.  
By applying mutation operators more than once  we obtain higher-order mutants. Higher-order mutants can also be described as a combination of several first-order mutants. 
Jia et al.~\cite{Jia2008} introduced the concept of higher-order mutation testing and discussed the relation between higher-order mutants and first-order mutants. 

They divided the higher-order mutants into four categories based on the observed coupling effect~\cite{Jia2009}: \textit{Expected}, \textit{Worst}, 
\textit{Fault Shift}, and \textit{Fault Mask}.
However, this categorization cannot be used to clarify the \textit{unexpected status} of the higher-order mutants, mainly because it does not consider how the status of the higher-order mutant relates to the status of the underlying first-order mutants. Therefore, we propose the following categorization:
\begin{itemize}
	\item \textbf{Expected.} The higher-order mutant is killed, and at least one of the underlying first-order mutants is killed; or the higher-order mutant has survived as well as all the underlying first-order mutants.
	
	\item \textbf{HK-FS.} The higher-order mutant is killed, even though all the underlying first-order mutants have survived.

	\item \textbf{HS-FK.} The higher-order mutant has survived, even though at least one of the underlying first-order mutants is killed.

\end{itemize}

When considering the overall percentage for each class, the mutants in categories HK-FS and HS-FK compensate the effects of each other. 
Therefore, if there are the same number of mutants in each of these categories, it is not visible in the final mutation coverage. 
The model must take into account these interactions among categories to correctly describe the real behavior of higher-order mutants.

\subsection*{LittleDarwin}
To perform our analysis, we modified the LittleDarwin\footnote{\url{http://littledarwin.parsai.net/}} mutation testing tool previously used by Parsai et al.~\cite{Parsai2015,Parsai2015a,Parsai2016} to generate higher-order mutants. LittleDarwin creates mutants by manipulating source code, and keeps the information about generated mutants and the results of the analysis in a local database, allowing  to perform further analysis of the results.

In its current version, LittleDarwin supports mutation testing of Java programs with a total of 9 mutation operators. 
These mutation operators are an adaptation of the  minimal set introduced by Offutt et al.~\cite{Offutt1996}. 
The description of each mutation operator along with an example can be found in Table~\ref{mutationoperators}. 
LittleDarwin constructs higher-order mutants by combining two (or more) first-order mutants randomly at class-level. The information regarding the underlying first-order mutants are provided inside each higher-order mutant.
\begin{center}
	\begin{table}[!t]
		
		\centering
		\caption{LittleDarwin mutation operators}
		\label{mutationoperators}
		
		\adjustbox{max width=\linewidth}
		{\begin{tabular}{|l||l|c|c|}
				\hline \multirow{2}{*}{\textbf{Operator}} & \multirow{2}{*}{\textbf{Description}} & \multicolumn{2}{c|}{\textbf{Example}} \\
				\hhline{~~--} & & \textbf{Before} & \textbf{After} \\ 
				\hline
				\hline AOR-B & Replaces a binary arithmetic operator & $a + b$  & $a - b$ \\ 
				\hline AOR-S & Replaces a shortcut arithmetic operator & $++a$ & $--a$ \\ 
				\hline AOR-U & Replaces a unary arithmetic operator & $-a$ & $+a$ \\ 
				\hline LOR & Replaces a logical operator & $a\,\&\,b$ & $a\,|\,b$ \\ 
				\hline SOR & Replaces a shift operator & $a >> b$ & $a << b$ \\ 
				\hline ROR & Replaces a relational operator & $a >= b$ & $a < b$ \\ 
				\hline COR & Replaces a binary conditional operator & $a\:\&\&\:b$ & $a\,||\,b$ \\ 
				\hline COD & Removes a unary conditional operator & $!\,a$  & $a$ \\ 
				\hline SAOR & Replaces a shortcut assignment operator & $a\:*= b$ & $a\:/= b$ \\ 
				\hline 
			\end{tabular}}
						\end{table}
	\end{center}

\section{Proposed Model}
\label{sect:model}

In order to estimate the first-order mutation coverage from the higher-order mutation coverage (mutation coverage calculated using higher-order mutants)
we follow these steps:  \begin{itemize}
	\item We define a higher-order mutant as a combination of multiple first-order mutants: 
	\vspace{-0.5em}
	\begin{equation}
	 h = m_1, m_2, ... , m_n 
	 \end{equation}
	\vspace{-1em}
	
	\item We assume that the higher-order mutant is killed if and only if  	at least one of the underlying first-order mutants are killed (Expected category).  
	\\
The probability of a mutant ($P(mutant)$) being killed is defined as the likelihood of choosing a random mutant with a killed status out of all mutants.  
The probability of a higher-order mutant being killed ($P(h)$) can be calculated from the probability of underlying first-order mutants ($P(m_i)$) in the following manner: 

	\vspace{-0.5em}	
	\begin{eqnarray}
	1-P(h) &{} = {}& (1-P(m_1))  \\
	&&\times(1-P(m_2|m_1)) \nonumber \\
	&&... \nonumber \\
	&&\times (1-P(m_n|m_1 ,  m_2 , ... , m_{n-1}))\nonumber 
	\label{eq:probability1}
	\end{eqnarray}
	
	This simply means that the probability of a higher-order mutant being killed can be evaluated by an ordered evaluation of the probabilities of underlying first-order mutants being killed.
	    
	\item After calculating the higher-order mutation coverage, we assume all higher-order mutants have the same probability of being killed equal to this mutation coverage\footnote{For example, imagine a bag of colored balls, in which 30\% are red, and 70\% are blue. Now if a ball is picked at random, there is a 30\% probability that this ball is red.}.
	
	\item For the sake of estimation, we assume all first-order mutants have the same equal probability of being killed, and this  probability is equal to first-order mutation coverage. We assume these probabilities to be independent from each other.\\ 
	{\footnotesize$P(m_1)=P(m_2|m_1)=...=P(m_n|m_1 ,  m_2 , ... , m_{n-1})$}

	\end{itemize}

 To ease the calculation of mutation coverage, we simplify Equation~\ref{eq:probability1} using the above assumptions:
 
 \begin{equation}
 1-P(h) = (1-P(m))^n 
 \label{eq:probability2}
 \end{equation}

Using  Equation~\ref{eq:probability2}, we assume $P(h)$ equals to the higher-order mutation coverage, i.e. any selected higher-order mutant from a set containing $n\%$ killed higher-order mutants is a killed one with the probability of $n\%$.  From this equation, we derive the value for $P(m)$, which is the probability of a single first-order mutant being killed, and we use this value as our estimation for first-order mutation coverage (Equation~\ref{eq:probability3}).  
 
\begin{equation}
P(m) = 1 - \sqrt[n]{1-P(h)}
\label{eq:probability3}
\end{equation}

\begin{figure}[!t]
	\centering
	\includegraphics[width=0.7\linewidth]{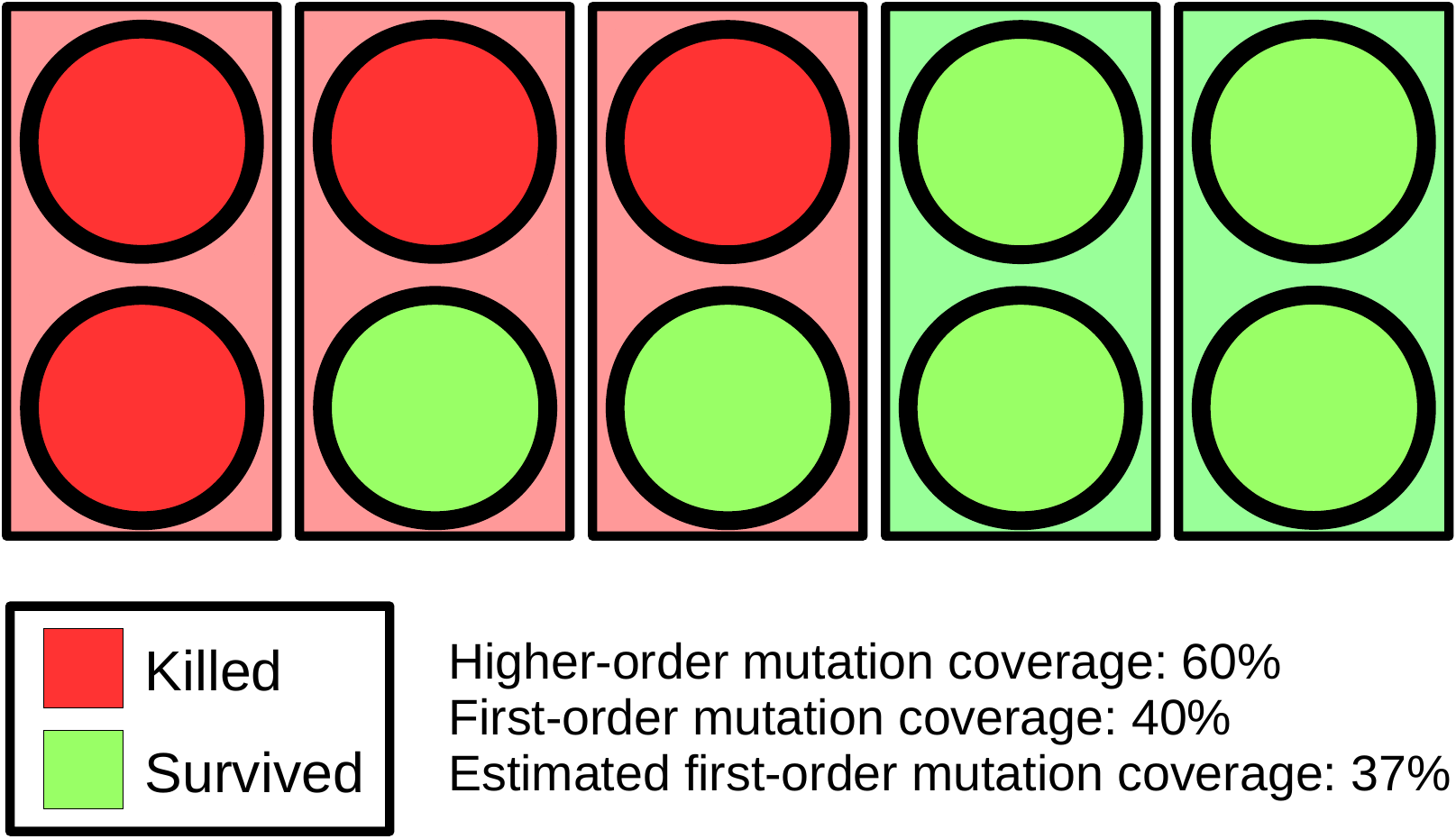}
	\caption{Example of the estimation provided by the model}
	\label{fig:methodexample}
\end{figure}

Figure~\ref{fig:methodexample} shows an example of this model; circles denote first-order mutants and rectangles are second-order mutants created by combining two first-order mutants. Out of 10 first-order mutants, 4 are killed, therefore the mutation coverage for this example is equal to 40\%. However, 3 out of 5 higher-order mutants are killed, resulting in 60\% coverage calculated using higher-order mutants. Considering our model, we estimate the first-order mutation coverage using Equation~\ref{eq:probability3} as 37\%.

\section{Case Study Design}
\label{sect:csdesign}

In order to evaluate our model, we performed first-, second-, third-, fourth-, fifth-, sixth-, and eighth-order mutation testing on 17 different Java projects. For these projects, 20849 first-order mutants were generated for a total of 1022 classes.
In this section, we first describe our cases, and then we discuss how the study was performed.

\subsection*{Cases}
We selected 17 open-source projects for our empirical study (Table~\ref{table:cases}). These projects are 
Joda Time\footnote{\url{http://www.joda.org/joda-time/}}, 
Apache Commons Lang\footnote{\url{http://commons.apache.org/proper/commons-lang/}}, 
Apache Commons Codec\footnote{\url{http://commons.apache.org/proper/commons-codec/}}, 
AddThis Codec\footnote{\url{http://github.com/addthis/codec}}, 
Apache Commons CLI\footnote{\url{http://commons.apache.org/proper/commons-cli/}}, 
Apache Commons FileUpload\footnote{\url{http://commons.apache.org/proper/commons-fileupload/}},
JSQLParser\footnote{\url{https://github.com/JSQLParser/JSqlParser}}, JDepend\footnote{\url{http://www.clarkware.com/software/JDepend.html}}, JGraphT\footnote{\url{http://jgrapht.org/}},  
JTerminal\footnote{\url{https://grahamedgecombe.com/projects/jterminal}},   
VRaptor\footnote{\url{http://www.vraptor.org/}}, 
PITest\footnote{\url{http://pitest.org/}}, 
HTTP Request\footnote{\url{http://kevinsawicki.github.io/http-request/}},
jsoup\footnote{\url{http://jsoup.org/}},
Linq4J\footnote{\url{http://www.hydromatic.net/linq4j}},
ScribeJava\footnote{\url{https://github.com/scribejava/scribejava}},
and jOpt Simple\footnote{\url{http://pholser.github.io/jopt-simple/}}.

The selected projects differ in size of their production code, test code, number of commits, and team size  to provide a wide range of possible scenarios. Moreover, they also differ in adequacy of the test suite.
based on statement, branch, and mutation coverage (Table~\ref{table:cases}). All selected projects are written in Java, which is a widely used programming language in industry~\cite{Garousi2013}.
\begin{table}
	\centering
	\caption{Projects sorted by mutation coverage}
	\label{table:cases}
	
	\adjustbox{max width=\linewidth}{
		\begin{tabular}{|c|c|c|c|c|c|c|c|c|c|}
			\hline \multirow{2}{*}{\textbf{Project}} & \multirow{2}{*}{\textbf{Ver.}} & \multicolumn{2}{c|}{\textbf{Size (LoC)}} & \multirow{2}{*}{\textbf{\#C}} & \multirow{2}{*}{\textbf{TS}} & \multirow{2}{*}{\textbf{SC}} & \multirow{2}{*}{\textbf{BC}} & \multirow{2}{*}{\textbf{MC}} & \multirow{2}{*}{\textbf{\#M}}\\
			\hhline{~~--~~~~} &  & \textbf{Prod.} & \textbf{Test} &  &  & & & & \\ 
			\hline
			\hline ScribeJava & 1.3.0 & 2002 & 1530   & 525 & 69 & 43\% & 66\% & 94.9\% & 59\\ 
			\hline Apache Commons CLI & 1.3.1   & 2665 & 3768  & 816 & 15 & 96\% & 93\% & 94.2\% & 342\\ 
			\hline JSQLParser & 0.9.4 & 7342 & 5909  & 576 & 19 & 81\% & 73\% & 93.6\% & 488\\ 
			\hline jOpt Simple & 4.8 &  1982 & 6084 & 297 & 14 & 99\% & 97\% & 91.7\% & 206\\ 
			\hline Apache Commons Lang & 3.4   & 24289 & 41758 & 4398 & 30 & 94\% & 90\% & 90.7\% & 6014 \\ 
			\hline Joda Time & 2.8.1  & 28479 & 54645 & 1909 & 42 & 90\% & 81\% & 81.7\% & 4870\\ 
			\hline Apache Commons Codec & 1.10   & 6485 & 10782 & 1461 & 10  & 96\% & 92\% & 81.6\% & 1976 \\ 
			\hline VRaptor & 3.5.5 & 14111 & 15496  & 3417& 65 & 87\% & 81\% & 81.2\% & 589\\ 
			\hline HTTP Request & 6.0 & 1391 & 2721   & 446 & 15 & 94\% & 75\% & 78.4\% & 227 \\ 
			\hline Apache Commons FileUpload & 1.3.1 & 2408 & 1892   & 846 & 19 & 76\% & 74\% & 77.1\% & 354\\ 
			\hline jsoup & 1.8.3 & 10295 & 4538  & 888 & 43 & 82\% & 72\% & 76.1\% & 1219\\ 
			\hline JGraphT & 0.9.1 & 13822 & 8180  & 1150& 31 & 79\% & 73\% & 69.4\% & 1356\\ 
			\hline AddThis Codec & 3.3.0 & 3675 & 1342  & 249& 4 & 69\% & 63\% & 64.7\% & 450\\ 
			\hline PITest & 1.1.7 & 17244 & 19005  & 1044 & 19 &  79\% & 73\%  & 62.9\% & 1070\\ 
			\hline JTerminal  & 1.0.1 & 687 & 250  & 8 & 2 & 66\% & 56\% & 60.0\% & 160\\ 
			\hline JDepend & 2.9.1 & 2460 & 1053   & 18 & 2 & 59\% & 52\% & 59.0\% & 239\\ 
			\hline Linq4J & 0.4 & 14307 & 3979   & 205 & 7 & 33\% & 66\% & 46.2\% & 1230\\ 
			\hline
						\multicolumn{10}{c}{} \\
			\multicolumn{1}{l}{Acronyms:} & \multicolumn{9}{r}{Version (Ver.), Line of code (LoC), Production code (Prod.),} \\
			\multicolumn{10}{r}{Number of commits (\#C),  Team size (TS), Statement coverage (SC),}\\
			\multicolumn{10}{r}{Branch coverage (BC), Mutation coverage (MC), Number of Mutants (\#M)}
		\end{tabular}

	}	
\end{table}

\subsection*{Model Evaluation}

\begin{figure}
\centering
\includegraphics[width=\linewidth]{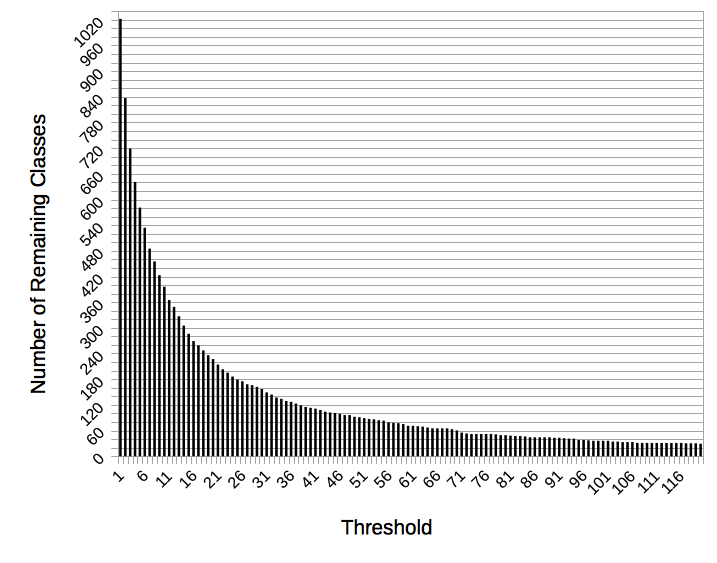}
\caption{Number of remaining classes after applying threshold $t$}
\label{fig:threshold-numberofclasses}
\end{figure}

The accuracy of the model is affected by the accuracy of $P(m)$. 
This in turn depends on the number of mutants  generated for the class in question (10 mutants would result in a probability with 0.1 accuracy, while 100 mutants improves this to 0.01). 
We define a threshold $t$ as the  number of minimum generated mutants for a class, and  we filter out the classes with less than $t$ generated mutants. 
The higher the threshold, the fewer the classes considered. This can be seen in  Figure~\ref{fig:threshold-numberofclasses}, where the x axis is the threshold, and the y axis is the number of remaining classes after applying the threshold.
In our experiment, we decided to evaluate how the model behaves with different values of $t$ in order to find a value that still keeps as many classes as possible, while filtering out less accurate data. 

The first goal of our work is to find a model that estimates ---as best as possible--- the first-order mutation coverage using higher-order mutants coverage. 
The second one is to create a model able to justify the obtained estimation.
For the first goal, a simple polynomial regression analysis is used in order to find the best fitting curve of the empirical data.
The order of the polynomial function influences the quality of the fitting: the higher the order, the higher the chance that the curve (over)fits the all points. 
However, to satisfy the second goal, we also need to find a curve that describes the \textit{real} behavior of the empirical data. 
For this reason, we cannot use a polynomial fitting curve of third or higher order. 
A  polynomial curve of third order has 1 ($3-2$) inflections, namely there exist a range of values where an increment of the first-order coverage
 leads to a decrement of the second-order coverage. 
The latter event cannot happen since it would imply that the more first-order mutants are killed, the less higher-order mutant are killed.
In our experiment, we compare the best second order fitting curve with the one provide by our model. This allows us to assess how far our model is from the optimal estimation.

\section{Results and Discussion}
\label{sect:results}
In this section, for each research question, we first discuss the motivation, then we explain our approach, and finally we present our findings. 

\subsection*{\textit{RQ1. \RQA}} 

\textbf{Motivation.} We are interested to verify whether the higher-order 
mutants can be used in place of the  first-order mutants ---with negligible drawbacks--- for the estimation of the quality of the test suite. 
For the sake of clarity, this RQ analyzes the model based only on second-order mutants. 
If with this order the model performs poorly, then there would be no reason to extend the analysis to higher orders.
We want to verify the applicability of the model, and for this reason, we inspect two aspects: 
\begin{itemize}
\item  \textit{Estimation Accuracy}. 
	The second-order mutation coverage overestimates the capabilities of the test suite, i.e. second-order mutants are easier to kill compared to first-order mutants. 
	We need to empirically evaluate the accuracy of the estimations based on second-order mutation coverage.
\item  \textit{Soundness of the Model}. 
	Our model is based on the assumption that second-order mutants are killed \textit{if and only if} at least one of the underlying first-order mutants is killed.
	We need to verify if this assumption holds with the empirical data.
\end{itemize}
For this reason, we break the first research question in two parts:\\

\textit{a) \RQone} 

\noindent\textbf{Approach.} We validate our model using 17 open-source projects. 
For all projects, we perform both first-order and second-order mutation testing. 
Then, we verify how the threshold $t$ of the generated mutants affects the estimations. 
To evaluate the accuracy of the estimations, 
we calculate the coefficient of determination ($R^2$) between the results estimated by our model, and the empirical data~\cite[p. 78]{Everitt2002}.

We compare the curve predicted by our model with the best fitting second-order curve obtained by using polynomial regression analysis on the empirical data.
From this comparison, we can evaluate how close is the curve proposed by the model from the best possible one.

\begin{figure}
	\centering
	\includegraphics[width=\linewidth]{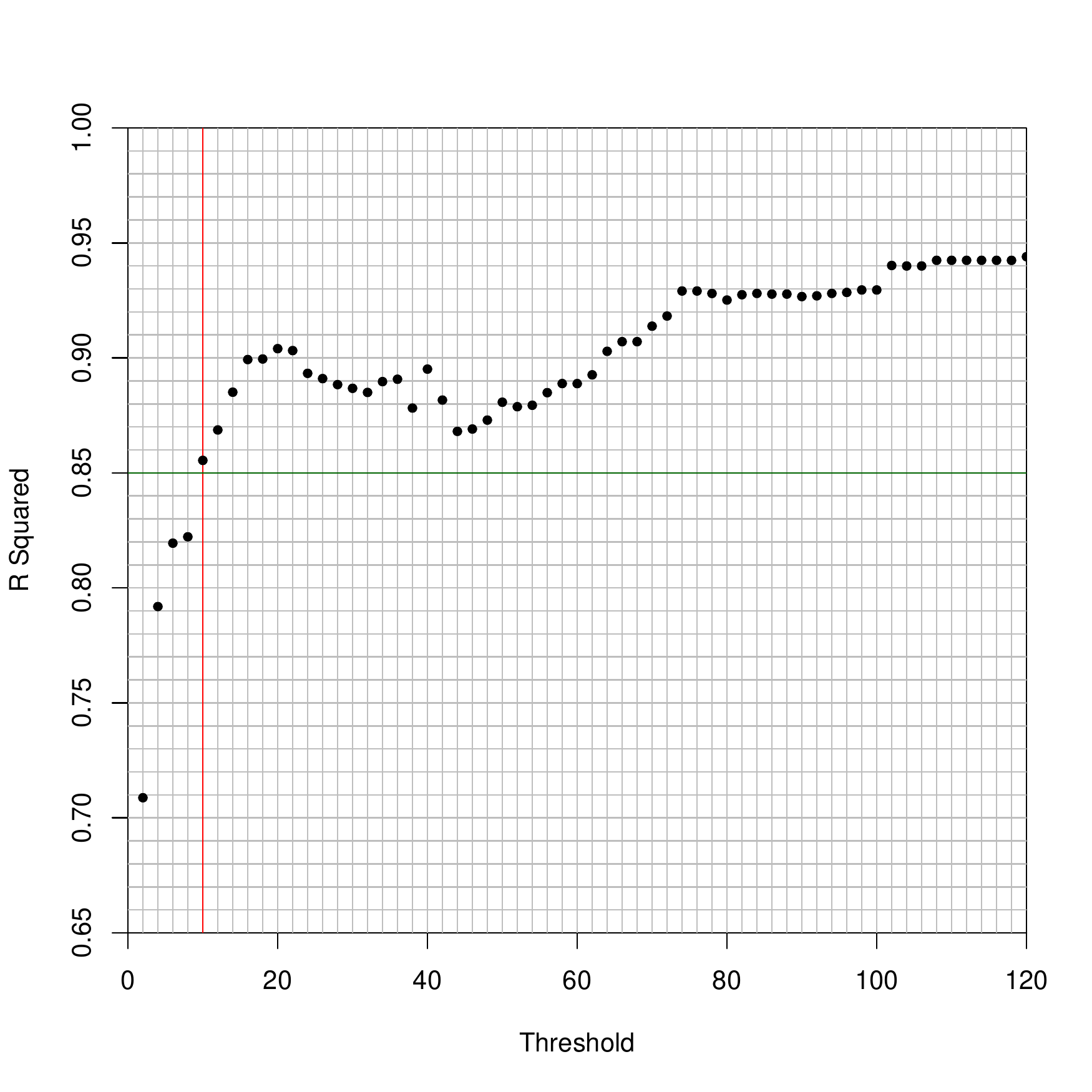}
	\caption{$R^2$ between estimated results and empirical data for different thresholds for second-order mutation. The red line denotes the chosen threshold, and the  green line shows the level of $R^2$ that the chosen threshold guaranties.}
	\label{fig:threshold}
\end{figure}

\textbf{Findings.} Figure~\ref{fig:threshold} shows the value of $R^2$ between estimated  results and empirical data for various values of the threshold  $t$. 
The estimations achieve an $R^2$ value  0.85 with a threshold $t$ as low as 10. 
Using such threshold, the model can still be applicable to 38.7\% of the classes which account for 89.8\% of the mutants.
For higher thresholds, the $R^2$ value is always higher than 0.85. These results show that the model can provide a good accuracy in estimating the first-order mutation coverage from the second-order mutation coverage. 
Yet, it halves the computation time required.

\begin{figure}
\centering
\includegraphics[width=\linewidth]{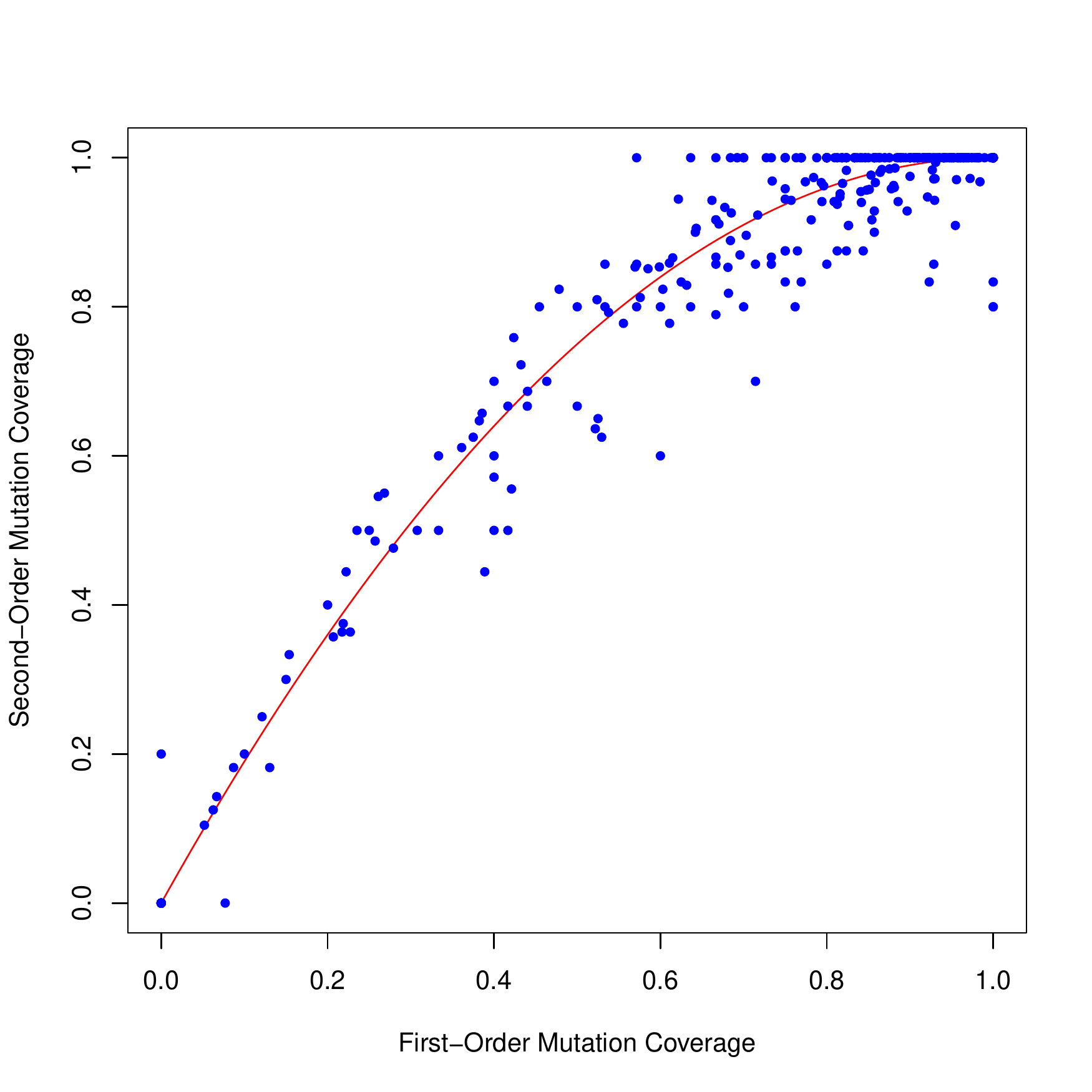}
\caption{Empirical data and the estimated curve for $t=10$. The red curve is the estimation provided by the model, and each blue dot represents a class.}
\label{fig:rq1}
\end{figure}

Figure~\ref{fig:rq1} shows the empirical data for $t=10$ and the estimated curve.
 The blue dots denote a class with an $x$ value of first-order mutation coverage and a $y$ value of second-order mutation coverage. 
 The red curve is the one predicted by our model: $y = -x^2 + 2x$.  
 The curve determined by the regression analysis is $y =  -1.008797x^2 + 1.999221x + 0.005342$. 
Table~\ref{table:parameters} shows the p-value for each estimated parameter. 
Considering the very low p-values calculated for each parameter, we can safely say that the trend of the empirical data is well represented by the best fitting curve.
As we can see, the curve predicted by our model, and the one provided by the regression analysis are very close in terms of coefficients. 
This highlights that our model provides (almost) the best possible estimations. The accuracy of the estimation, however, changes according to the mutation coverage value. This is observed especially for classes with a very high second-order mutation coverage, in which there is less information available for the model. At its extreme, when all the higher-order mutants are killed, the model always estimates a 100\% first-order coverage, even tough some of the underlying first-order mutants might survive.

\begin{table}
\centering

\caption{Parameters of the estimated curve}
\label{table:parameters}

	\adjustbox{max width=\linewidth}{
\begin{tabular}{|c|c|c|c|}

 \hline \textbf{Parameter} & \textbf{Estimated Value} & \textbf{Standard Error} &  \textbf{p-value} \\
     
 \hline\hline $x^0$  &  0.005342 &  0.010109  &   0.598 \\    
       \hline $x^1$  & 1.999221  &  0.037831  &  $<2\times10^{-16}$ \\
       \hline $x^2$  & -1.008797 & 0.032452 &  $<2\times10^{-16}$ \\
       \hline 
\end{tabular}
}
\end{table}

\textit{b) \RQtwo} 

\noindent\textbf{Approach.}
We categorize the second-order mutants into three categories: Expected, HK-FS, and HS-FK. 
Then we compute how many mutants are in each category.
The higher  the number in the Expected category, the better the model describes the behavior of the second-order mutants.

\textbf{Findings.} Table~\ref{table:rq2} shows the number of higher-order mutants in each category for each project. Out of 10153 second-order mutants generated for all projects, 10069 (99.17\%) are in the Expected category. This means that overall only 84 (0.83\%) mutants are of unexpected status, of which 13 (0.13\%) mutants belong 
to the HK-FS category, and 71 (0.70\%) mutants belong to the HS-FK category. In total, 764 (91.28\%) out of 837 classes do not contain any HS-FK or  HK-FS mutants. 
In almost all projects the number of mutants in the Expected category is higher than 97\%. 
The only exception is given by the project \textit{ScribeJava}. 
By manually checking this case, we found that
the unexpected second-order mutants were created in very small classes where the
 underlying first-order mutants directly interact with each other.     Overall, we see that the modeled behavior applies to the vast majority of the second-order mutants. 
\hypobox{ Our model  provides a good accuracy in estimating the first-order mutation coverage based on second-order mutation coverage. Moreover, it provides a curve which is
 close to the best one fitting the empirical data. 
 Finally, the behavior of the vast majority of the second-order mutants is modeled correctly.}

\begin{sidewaystable}
	\centering
	\caption{Number of higher-order mutants in each category for each project ($t=0$, all mutants included)}
	\label{table:rq2}

	\adjustbox{max width=\linewidth}{
		\begin{tabular}{|c|c|c|c|c|c|c|c|c|c|c|c|c|c|c|c|c|c|c|c|c|c|c|c|c|}
			
\hline \multirow{2}{*}{\textbf{Project}} & \multicolumn{4}{c|}{\textbf{Second-Order}} & \multicolumn{4}{c|}{\textbf{Third-Order}} & \multicolumn{4}{c|}{\textbf{Fourth-Order}} & \multicolumn{4}{c|}{\textbf{Fifth-Order}}& \multicolumn{4}{c|}{\textbf{Sixth-Order}}& \multicolumn{4}{c|}{\textbf{Eighth-Order}} \\
\hhline{~------------------------}  & \textbf{Exp. (\%)} & \textbf{HK-FS} & \textbf{HS-FK} & \textbf{Total} & \textbf{Exp. (\%)} & \textbf{HK-FS} & \textbf{HS-FK} & \textbf{Total} & \textbf{Exp. (\%)} & \textbf{HK-FS} & \textbf{HS-FK} & \textbf{Total} & \textbf{Exp. (\%)} & \textbf{HK-FS} & \textbf{HS-FK} & \textbf{Total}& \textbf{Exp. (\%)} & \textbf{HK-FS} & \textbf{HS-FK} & \textbf{Total}& \textbf{Exp. (\%)} & \textbf{HK-FS} & \textbf{HS-FK} & \textbf{Total} \\

\hline
\hline AddThis Codec  &  215 (99.08\%) &  1 &  1 &  217  &  135 (95.74\%)  &  1 &  5 &  141 & 98 (98.0\%) &  0 & 2  & 100 & 79 (98.75\%) &  0 & 1  & 80 & 64 (98.46\%) &  0 & 1  & 65 & 42 (95.45\%) &  0 & 2  & 44 \\
\hline Apache Commons CLI  &  164 (99.39\%) &  0 &  1 &  165  &  110 (100\%) &  0 &  0 &  110 & 81 (100.0\%) &  0 & 0  & 81 & 61 (100.0\%) &  0 & 0  & 61 & 52 (100.0\%) &  0 & 0  & 52 & 38 (100.0\%) &  0 & 0  & 38 \\
\hline Apache Commons Codec  &  978 (100\%) &  0 &  0 &  978 &  642 (100\%) &  0 &  0 &  642 & 479 (100.0\%) &  0 & 0  & 479  & 377 (100.0\%) &  0 & 0  & 377 & 309 (100.0\%) &  0 & 0  & 309 & 231 (100.0\%) &  0 & 0  & 231 \\
\hline Apache Commons FileUpload  &  174 (100\%) &  0 &  0 &  174  &  115 (100\%) &  0 &  0 &  115 	& 84 (100.0\%) &  0 & 0  & 84  & 56 (100.0\%) &  0 & 0  & 56 & 66 (100.0\%) &  0 & 0  & 66 & 38 (100.0\%) &  0 & 0  & 38 \\
\hline Apache Commons Lang  &  2976 (99.77\%) &  0 &  7 &  2983 &  1966 (99.95\%) &  0 &  1 &  1967 & 1468 (99.86\%) &  0 & 2  & 1470  & 1164 (100.0\%) &  0 & 0  & 1164 & 960 (100.0\%) &  0 & 0  & 960 & 714 (100.0\%) &  0 & 0  & 714\\
\hline HTTP-Request  &  113 (100\%) &  0 &  0 &  113 &   75 (100\%) &  0 &  0 &  75 & 56 (100.0\%) &  0 & 0  & 56  & 45 (100.0\%) &  0 & 0  & 45 & 37 (100.0\%) &  0 & 0  & 37 & 28 (100.0\%) &  0 & 0  & 28\\
\hline JDepend  &  111 (97.37\%) &  2 &  1 &  114  &  73 (98.65\%) &  1 &  0 &  74 	& 52 (100.0\%) &  0 & 0  & 52 & 42 (100.0\%) &  0 & 0  & 42 & 28 (84.85\%) &  0 & 5  & 33 & 23 (100.0\%) &  0 & 0  & 23\\
\hline JGraphT  &  638 (97.70\%) &  2 &  13 &  653 &  410 (99.03\%) &  1 &  3 &  414 & 290 (97.97\%) &  2 & 4  & 296 & 223 (98.67\%) &  1 & 2  & 226 & 182 (98.38\%) &  1 & 2  & 185 & 123 (99.19\%) &  0 & 1  & 124\\
\hline Joda Time  &  2388 (99.33\%) &  4 &  12 &  2404  &  1572 (99.49\%) &  3 &  5 &  1580 & 1169 (99.91\%) &  0 & 1  & 1170  & 919 (99.78\%) &  1 & 1  & 921 & 759 (100.0\%) &  0 & 0  & 759 & 556 (100.0\%) &  0 & 0  & 556\\
\hline jOpt Simple  &  90 (98.9\%) &  0 &  1 &  91 &  56 (100\%) &  0 &  0 &  56 & 41 (100.0\%) &  0 & 0  & 41 & 33 (100.0\%) &  0 & 0  & 33 & 24 (100.0\%) &  0 & 0  & 24 & 16 (100.0\%) &  0 & 0  & 16 \\
\hline jsoup &  596 (99.33\%) &  0 &  4 &  600 &  392 (100\%) &  0 &  0 &  392 & 288 (99.31\%) &  2 & 0  & 290 & 225 (99.12\%) &  2 & 0  & 227 & 189 (100.0\%) &  0 & 0  & 189 & 133 (100.0\%) &  0 & 0  & 133 \\
\hline JSQL Parser  &  213 (97.71\%) &  0 &  5 &  218 &  134  (100\%) &  0 &  0 &  134  & 84 (100.0\%) &  0 & 0  & 84 & 64 (100.0\%) &  0 & 0  & 64 & 47 (100.0\%) &  0 & 0  & 47 & 26 (100.0\%) &  0 & 0  & 26 \\
\hline JTerminal  &  79 (100\%) &  0 &  0 &  79  &  50 (98.04\%) &  0 &  1 &  51 & 37 (100.0\%) &  0 & 0  & 37 & 29 (100.0\%) &  0 & 0  & 29 & 24 (100.0\%) &  0 & 0  & 24 & 17 (100.0\%) &  0 & 0  & 17 \\
\hline Linq4J  &  594  (98.67\%) &  2 &  6 &  602  &  384 (99.22\%) &  1 &  2 &  387 & 285 (99.65\%) &  0 & 1  & 286 & 220 (98.65\%) &  0 & 3  & 223 & 176 (98.32\%) &  0 & 3  & 179 & 127 (99.22\%) &  0 & 1  & 128 \\
\hline PITest  &  475 (97.94\%) &  1 &  9 &  485  &  291 (97.65\%) &  2 &  5 &  298 & 198 (97.06\%) &  0 & 6  & 204 & 147 (95.45\%) &  1 & 6  & 154& 114 (97.44\%) &  1 & 2  & 117 & 71 (95.95\%) &  2 & 1  & 74 \\
\hline ScribeJava  &  20 (76.92\%) &  0 &  6 &  26  &  13 (100\%) &  0 &  0 &  13 & 9 (100.0\%) &  0 & 0  & 9 & 5 (100.0\%) &  0 & 0  & 5 & 4 (100.0\%) &  0 & 0  & 4 & 3 (100.0\%) &  0 & 0  & 3 \\
\hline VRaptor  &  245 (97.61\%) &  1 &  5 &  251  &  140 (99.29\%) &  1 &  0 &  141 & 91 (98.91\%) &  1 & 0  & 92  & 61 (95.31\%) &  1 & 2  & 64 & 48 (97.96\%) &  1 & 0  & 49 & 23 (95.83\%) &  1 & 0  & 24 \\
    			
\hline \hline Total & 10069 (99.17\%) & 13 & 71 & 10153  & 6459 (99.51\%) & 10 & 22 & 6590 & 4810 (99.57\%) &  5 & 16  & 4831 & 3760 (99.44\%) &  6 & 15  & 3781 & 3073 (99.48\%) &  3 & 13  & 3089 & 2209 (99.64\%) &  3 & 5  & 2217 \\
\hline
\end{tabular} }
	
\end{sidewaystable}

\subsection*{\textit{RQ2. \RQthree}} 
\textbf{Motivation.} 
RQ1 provides evidence of the soundness of the model and shows that second-order mutants can be used by the model to provide good estimations. 
In this RQ, we are interested in extending this analysis to verify how the higher orders of mutation affects the estimations.
Taking into account the trade-off between accuracy of the estimation and computational time required, this RQ helps testers to decide which order of mutation better fits their needs.

\textbf{Approach.} 
For all projects, we perform third-, fourth-, fifth-, sixth-, and eighth-order mutation testing.  
Then, we observe how the order of mutation affects the accuracy of the estimations for various thresholds. 
We also  verify whether the model describes the real behavior by computing the number of mutants in the Expected category for the aforementioned orders of mutation.

\begin{figure}
	\centering
	\begin{minipage}{0.9\linewidth}
	\includegraphics[width=\linewidth]{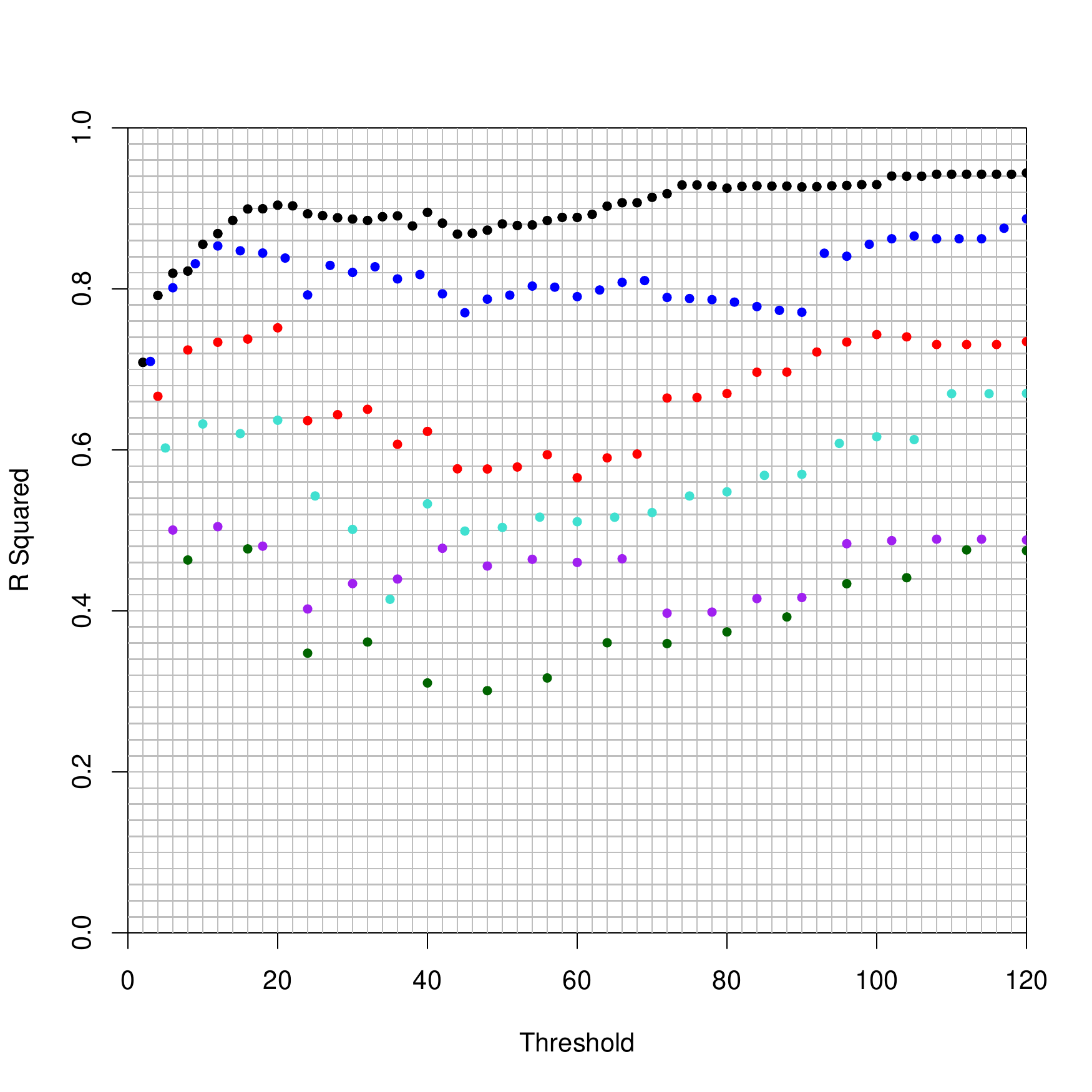}
	\end{minipage}\begin{minipage}{0.1\linewidth}
		\includegraphics[width=\linewidth]{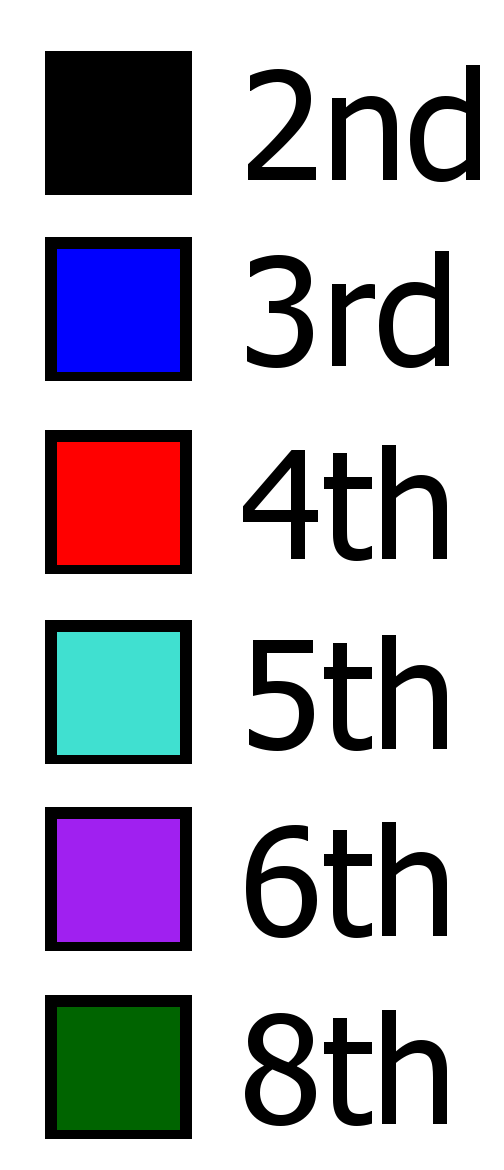}
	\end{minipage}

	\caption{$R^2$ between estimated results and empirical data for different thresholds.}
	\label{fig:thresholdall}
\end{figure}

\textbf{Findings.}
Figure~\ref{fig:thresholdall} shows the computed $R^2$ between estimated results and empirical data for various thresholds $t$ for second, third, fourth, fifth, sixth and eighth orders of mutation. In this figure, we observe that with the increment of the order of mutation, the  $R^2$ values decrease for any threshold. 
This can be attributed to the fact that the higher the order of the mutant is, the higher is the chance that at least one of the underlying first-order mutants is killed. 
This means that increasing the order of mutation makes the model less accurate. 
From the figure we notice that  the values of $R^2$ remain similar for the second-order and the third-order within $2 \leq t \leq  12$. 
Whereas, the values of $R^2$ for fourth and higher orders are visibly lower,  highlighting the limitation of proposed model.

As seen in Table~\ref{table:rq2}, the percentage of the mutants in the Expected category generally increases along with the order of mutation. 
This is in accordance to the results of the previous research~\cite{Offutt1992,Wah2003} on coupling effect, confirming the fact that the higher the order of mutation, the lesser the  chance of unexpected behavior. The same effect can symmetrically be  seen focusing on the decreasing number of mutants in the categories HS-FK and HK-FS. 
For HS-FK, there is lesser chance of survival for the higher-order mutant as the order of mutation increases. 
For HK-FS, there is lesser chance that all the underlying first-order mutants have survived when creating higher-order mutants randomly.

\hypobox{ 	The estimations of the model have an accuracy that decreases with the increment of the order of mutation. This decay is prominent from the fourth order of mutation onwards.
	 	}

\section{Threats to Validity}
\label{sect:threats}

To describe the threats to validity we refer to the guidelines reported by Yin~\cite{Yin2003}. 

Threats to \textbf{internal validity} focus on confounding factors that can influence the obtained results. 
This threat stems from potential bugs hidden inside the algorithms used for creating higher-order mutants or in LittleDarwin. 
We consider this chance ---even if possible--- limited. The results of the experiments are available, and several iterations of post-analysis were performed to make sure of the correctness of the original algorithms.
In addition, the code of LittleDarwin has been already checked and tested in several case studies~\cite{Parsai2015,Parsai2015a,Parsai2016}.

Threats to \textbf{external validity} refers to the generalizability of our results. In our analysis we use only 17 open-source projects. We mitigate this threat
by using projects  which differ for number of contributors, size and adequacy of the test suite. Yet,  it is desirable to replicate this study using more projects, 
especially the ones belonging to industrial settings. Another threat to external validity arises from the overfitting of the model to the data. Since our model is developed independent of the subjects of the study, we believe this threat does not apply.  

Threats to \textbf{construct validity}  are concerned with how accurately the observations describe the phenomena of interest. In our case, the techniques adopted to evaluate the accuracy of the estimations are: (i) calculation of coefficient of determination ($R^2$) between the estimated results and empirical data, and (ii) polynomial regression analysis. Both of these  techniques are  well known, and they have been used in literature numerous times for similar purposes. 
Another threat to construct validity stems from the fact that the process of creating higher-order mutants requires to combine several first-order mutants selected at random within a class. Therefore, multiple analysis would be needed to reduce the random effect. However, this is not possible, since mutation testing is a time-consuming procedure, and it is not possible to perform all experiments several times. Using $R^2$ for a data set of 837 extracted classes alleviates this issue to a certain extent by reducing the random effect over the whole dataset. We also use multiple cut-off thresholds to remove outliers from the dataset.    

The quality of the mutants affects the results of our study in two ways:
First, existence of equivalent first-order mutants reduces the accuracy of the first-order mutation coverage. However, we did not filter for the equivalent mutants, since when combined with a non-equivalent mutant, they do not affect the status of  the created  higher-order mutant. In other words, the higher-order mutant created from the combination of a killed first-order mutant and an equivalent one is still killed. The chance of a higher-order mutant consisting only out of equivalent mutants decreases exponentially by the order of mutation\footnote{For example, if 10\% of the first-order mutants are equivalent, the chance of a second-order mutant created by the combination of two equivalent mutants is close to 1\%.}. 
Second, Amman et al.~\cite{Ammann2014} show that large portions of first-order mutants are redundant for all practical purposes. This means that without adequate filtering of the first-order mutants, the resulting mutation coverage does not measure the quality of the test-suite accurately. Since this threat affects both first-order and second-order mutants, the extent of the problem for our empirical validation remains unknown, and requires further investigation.

\section{Related Works}
\label{sect:rw}
Reduction of the number of mutants has been investigated in the literature to reduce the computational cost of mutation testing. 
Higher-order mutation is first introduced by Offutt in 1992~\cite{Offutt1992} to investigate the coupling effect empirically. Jia and Harman provide a technique to create ``hard to kill'' higher-order mutants~\cite{Jia2009}. 
Polo et al.~\cite{Polo2009} and Papadakis et al.~\cite{Papadakis2010} evaluated second-order mutation testing by combining first-order mutants using different algorithms and concluded that second-order mutation testing reduces the effort significantly while also reducing the number of equivalent mutants. 

Differently from
these studies, our study focuses on the accuracy to the estimation of mutation coverage rather than on the creation of hard to kill mutants.
In contrast to previous studies where such mutants were desirable, in our case they represent a problem since they make the estimation of first-order mutation coverage less accurate in our model (we assume the faults behave as expected by the coupling effect when combined). 
In this sense, our work is more geared towards development environments that require an accurate metric to quantify the quality of their test suites.

Kintis et al.~\cite{Kintis2010} shows that using only disjoint second-order mutants creates a more robust measure of test suite effectiveness than first-order mutation coverage. Our model can be used in combination with their method of creating second-order mutants, however, an empirical evaluation of such combination remains as future work.

There are several studies that attempt to validate the coupling effect hypothesis both theoretically and empirically. 
Offutt provided experiments in support of the hypothesis showing that the vast majority of the higher-order mutants are coupled to the first-order mutants ~\cite{Offutt1989,Offutt1992}. 
Wah designed a mathematical model to describe the behavior of the faults in a program consisting only of functions~\cite{Wah2000,Wah2003}. 
He showed that by increasing the order of the mutation, the number of de-coupled mutants decrease.
 Our study agrees with the aforementioned studies, as we show that the number of higher-order mutants with unexpected behavior is very small, and decreases when the order of mutation increases.

\section{Conclusion}
\label{sect:conclusion}
First-order mutation coverage is an effective metric to evaluate the quality of the test suite. However, its adoption is hindered due to the high computational time consumption.
On one hand,  the adoption of higher-order mutants to evaluate the quality of the test suite is a viable solution. 
On the other hand, this approach provides less realistic estimations than the ones obtained with first-order mutants. 
As a solution, we propose a realistic model able to estimate first-order mutation coverage based on higher-order mutation coverage. 
The benefits of this model are that (i) we achieve a good accuracy in estimating the first-order mutation coverage based on second- and third-order mutation coverage,
 (ii) we halve the computational time that would be required by first-order mutation testing (at minimum). 
 Moreover, the chance of existence of equivalent mutants is less than first-order analysis, even though the model does not explicitly consider them in its estimation.  Finally, the model correctly describes the real behavior of vast majority of higher-order mutants.

\section*{Acknowledgments}
This work is sponsored by the Institute for the Promotion of Innovation through Science and Technology in Flanders through a project entitled Change-centric Quality Assurance (CHAQ) with number 120028.

\balance
\printbibliography

\end{document}